\documentclass[a4paper]{aa}
\usepackage{amsmath}
\usepackage{txfonts}
\usepackage{graphicx}
\usepackage{natbib}
\usepackage{multirow}
\usepackage{color}
\usepackage{array}
\usepackage{ifthen, hyperref}
\usepackage{fixltx2e}
\usepackage{caption}
\usepackage{lscape}
\usepackage{tabularx}
\usepackage{breakurl}
\usepackage{enumitem}
\bibpunct{(}{)}{;}{a}{}{,}

\begin{document}

\title{A multi-wavelength approach to classifying transient events in the direction of M31}
\author{Monika~D.~Soraisam\thanks{soraisam@noao.edu} \inst{\ref{NOAO}}
    \and Marat~Gilfanov \inst{\ref{MPA},\ref{SRI}}
    \and Thomas~Kupfer \inst{\ref{Caltech}}
    \and Thomas~A.~Prince\inst{\ref{Caltech}}
    \and Frank~Masci\inst{\ref{IPAC}}
    \and Russ~R.~Laher\inst{\ref{IPAC}}
    \and Albert~K.~H.~Kong\inst{\ref{NTHU}}
    }
\institute{National Optical Astronomy Observatory, Tucson, AZ 85719, USA\label{NOAO}
  \and Max Planck Institute for Astrophysics, Karl-Schwarzschild-Str.~1, 85748 Garching, Germany\label{MPA}
  \and Space Research Institute, Russian Academy of Sciences, Profsoyuznaya~84/32, 117997 Moscow, Russia\label{SRI}
  \and Division of Physics, Mathematics, and Astronomy, California Institute of Technology, Pasadena, CA 91125, USA\label{Caltech}
  \and Infrared Processing and Analysis Center, California Institute of Technology, Pasadena, CA 91125, USA\label{IPAC}
  \and Institute of Astronomy, National Tsing Hua University, Hsinchu 30013, Taiwan\label{NTHU} 
    }

\date{Received date / Accepted date}

\abstract
{In the hunt for rare time-domain events, confusing exotic extragalactic phenomena with more common Galactic foreground events is an important consideration.}
{We show how observations from multiple wavebands, in this case optical and X-ray observations, can be used to facilitate the distinction between the two.}
{
We discovered an extremely bright and rapid transient event during optical observations of the M31 galaxy  taken by the intermediate Palomar Transient Factory (iPTF). The persistent optical counterpart  of this transient  was previously thought to be a variable star in M31 without any dramatic flux excursions. The iPTF event initially appeared to be an extraordinarily rapid and energetic extragalactic transient, a very fast  nova in M31 being one of the exciting possibilities, with a $\approx 3$~mag positive flux excursion in less than a kilosecond. The nature of the source was resolved with the help of Chandra archival data, where we found an X-ray counterpart and obtained its X-ray spectrum.
}
{
We find the X-ray spectrum of the quiescent emission can be described by a model of optically thin plasma emission with a temperature of $\approx 7$~MK, typical for coronal emission from an active star. The combination of the X-ray  luminosity, calculated assuming the source is located in M31 ($\sim 3\times10^{36}$~erg/s), and the  color temperature exclude any type of known accreting compact object or active star in M31. We argue instead that the optical transient source is an M type main-sequence, active star located in the disk of the Milky Way at a distance of $\sim 0.5\mbox{--}1$~kpc. Its persistent X-ray luminosity is in the $\approx 1.3\mbox{--}5\times10^{30}$~erg/s range and it has the absolute optical magnitude of 9.5 --11.0~mag in R-band. The observed optical flare has the equivalent duration of $\approx 95$ min and total energy of $\approx (0.3\mbox{--}1)\times10^{35}$~erg in R-band, which places it among the brightest flares ever observed from an M-type star. This case can serve as an example for the classification of Galactic and extragalactic events in upcoming high-cadence time-domain projects, such as the Zwicky Transient Facility and the Large Synoptic Survey Telescope.
}
{}
{}
\keywords{Stars: individual: WeCAPP V07979 -- galaxies: individual: M31 -- surveys -- X-rays.}

\authorrunning{Soraisam et al.}
\maketitle

\section{Introduction}\label{sec:intro}

In the last two decades, time-domain astronomy has experienced a period of massive acceleration, both in terms of the rate at which objects are discovered and in terms of the variability timescales of these objects. This development is due to a host of ground-based optical surveys. Particularly significant is the contribution of the (intermediate) Palomar Transient Factory (PTF/iPTF; \citealt{Law, Rau, Ofek-2012}) to the discovery of rapid transients, characterized by timescales on the order of days to weeks (see, e.g., \citealt{Kasliwal-2013}). The success of this transient factory is expected to expand even further with the upcoming Zwicky Transient Facility (ZTF; \citealt{Bellm-2014}).

An interesting endeavor in light of the flood of transient detections is to compile a complete catalog of the transient events within a particular galaxy. On the one hand, such a study would help to throw light on the properties of this galaxy, in particular its stellar population. For example, the underlying white dwarf (WD) population in cataclysmic variables could be determined using novae and dwarf novae. On the other hand, it would enable the identification of abnormal transients in future projects, such as ZTF and the Large Synoptic Survey Telescope (LSST), by comparison of the properties of an event to their expected distribution for this galaxy. To obtain such a catalog, we require a dedicated high-cadence monitoring of a galaxy, specifically one whose stars can be resolved as much as possible within the sensitivity of the observation system. The iPTF observations of M31, which we are using here, provide a combination of high cadence and a moderate angular resolution, thus providing a first step toward such a goal, at least for bright sources.

One problem, both in the creation of a catalog of transient events within a galaxy, and in the subsequent identification of exotic events, is the existence of foreground events. Objects residing within the Milky Way, in front of the target galaxy, are not always easily distinguishable from objects within the target galaxy itself. They will often exhibit properties that are uncharacteristic of extragalactic transient objects, such as the apparent magnitude of their brightness fluctuation. Example systems which have suffered confusion in their localizations include PT And and M31N~1966-08a (that spatially coincided with M31N~1968-10c). The outbursts from the former have been classified as recurrent nova explosions in M31 by some authors \citep[e.g.,][]{Cao-2012}, and by some as dwarf nova outbursts in our Galaxy, most recently by \citet{Williams-2017}. The second example too was initially considered to be a recurrent nova candidate in M31, but \citet{Shafter-2017RNAAS} has classified it as a flare star.
To avoid classifying events as rare occurrences of highly energetic extragalactic phenomena, it is important to devise alternative ways to distinguish exotic extragalactic events from more mundane Galactic transients.

A promising avenue for the distinction between moderate flux excursions in Galactic objects and extraordinarily large events in extragalactic objects is to utilize observations from multiple spectral domains. Here we use optical observations from the iPTF survey of one particular transient event in conjunction with X-ray observations from the \emph{Chandra} X-ray Observatory in the same direction to exemplify the detailed steps involved in such an approach and demonstrate its feasibility. We show how this event, which initially appeared to be an unusually fast and energetic transient in the galaxy M31, can be identified as a foreground flare star residing within the Milky Way, thanks to the multi-wavelength information. It should be noted though that in the coming era of data-deluge with \emph{wide-field} optical surveys such as ZTF and LSST, suitable successors to existing X-ray observatories---\emph{Chandra}, \emph{XMM-Newton}, \emph{Swift}---will be required.

The paper is organized as follows. In Sect.~\ref{sec:all_optical}, we describe the iPTF observations leading to the discovery of this transient, and the analysis of the optical (iPTF) data. In Sect.~\ref{sec:xray}, we present our analysis of the X-ray data, specifically from \textit{Chandra}. In Sect.~\ref{sec:result}, we examine the nature of the source in light of the observational results, and end with a discussion and conclusions in Sect.~\ref{sec:conclude}. 
We used a distance modulus of 24.36 for M31 \citep{Vilardell}.

\section{Optical data}\label{sec:all_optical}

\subsection{Optical time-domain data of M31}\label{sec:data_red}

We have started analyzing the iPTF observations of M31 with one of the motivations being to search for fast novae. These are transient events with decline times of hours to few days, and are expected to harbor massive WDs (e.g., \citealt{Prialnik-1995, Starrfield-2012, Kato-2014, Hillman-2016}). The importance of fast novae in the context of type Ia supernova progenitors has been extensively discussed by \citet{Soraisam-2015}. The task of searching for novae necessarily entails scanning through all possible transients that were detected by the survey. 
As such, there is also a scope of finding other, non-nova transients that exhibit characteristics of the sought lightcurve timescale, in particular a very rapid decline.      

The optical data analyzed here are of a single band, that is $R$ band, and were procured by the iPTF survey using the 48-inch Samuel Oschin Telescope, equipped with a detector covering $\approx 7.26$ square degrees with a pixel scale of $1.01''$. The depth of the survey reaches 21~mag in the $R$ band. The M31 field studied here is the same as the one studied by \citet{Soraisam-2017}, hereafter SG17. We have however extended the baseline to January 2013. 

The particular set of observations relevant for this paper are those of a specific night, namely those of 31 January 2013. This night has the highest number of visits among all the observed nights of M31 for which data are available to us. The total number of epochs on this night is 121, with a typical separation of around 2~minutes between consecutive observations. This night was an instance of the dynamic cadence experiment of iPTF aimed for exploring very fast transient phenomena on timescales between 1 minute and 5 days (see \citealt{Rau}).
In the following, all analyses refer to these 121 observation epochs and the source(s) obtained from them.

\begin{figure*}[t]
\centering
\includegraphics[width=88mm]{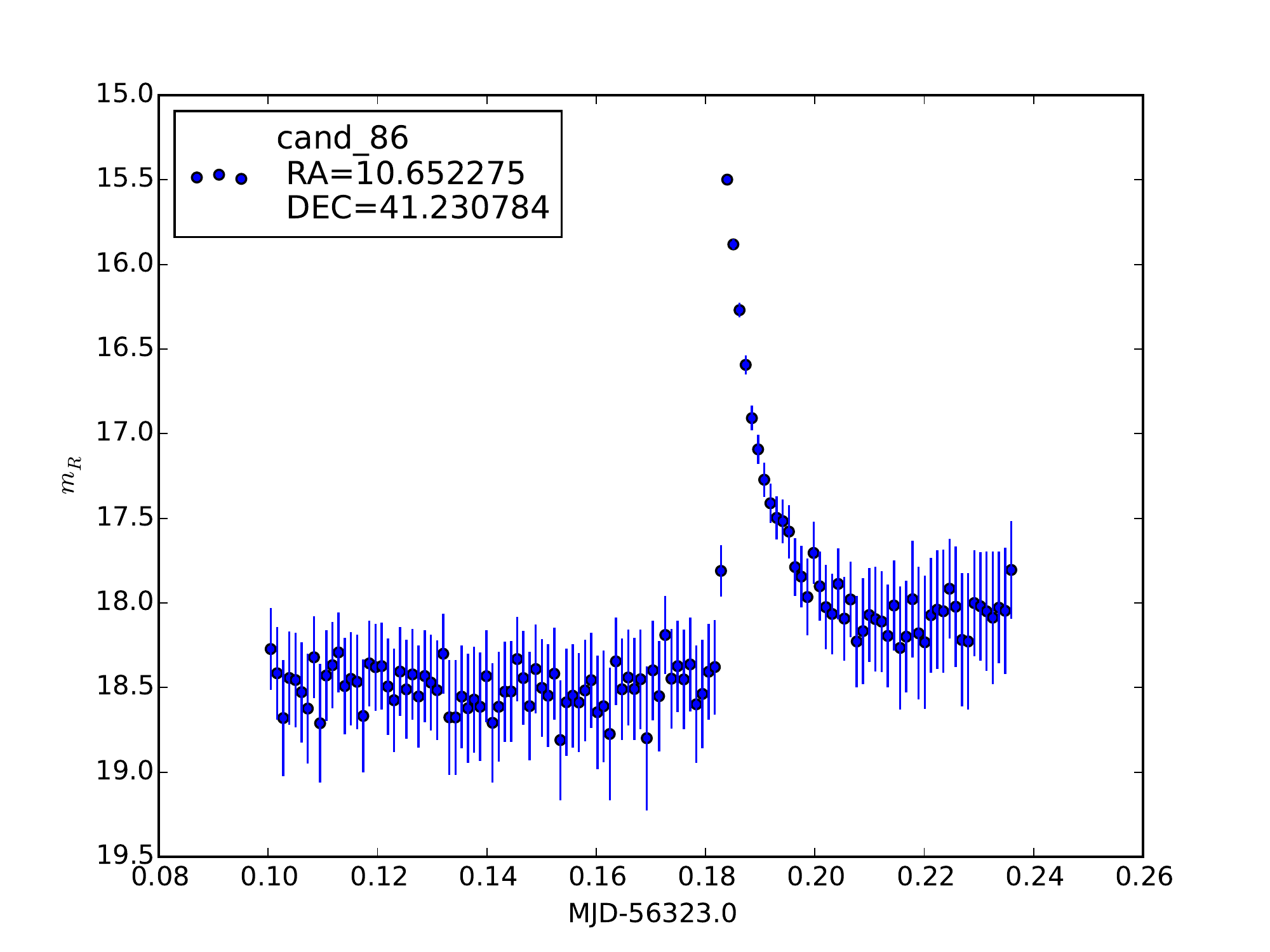}\hfill \includegraphics[width=88mm]{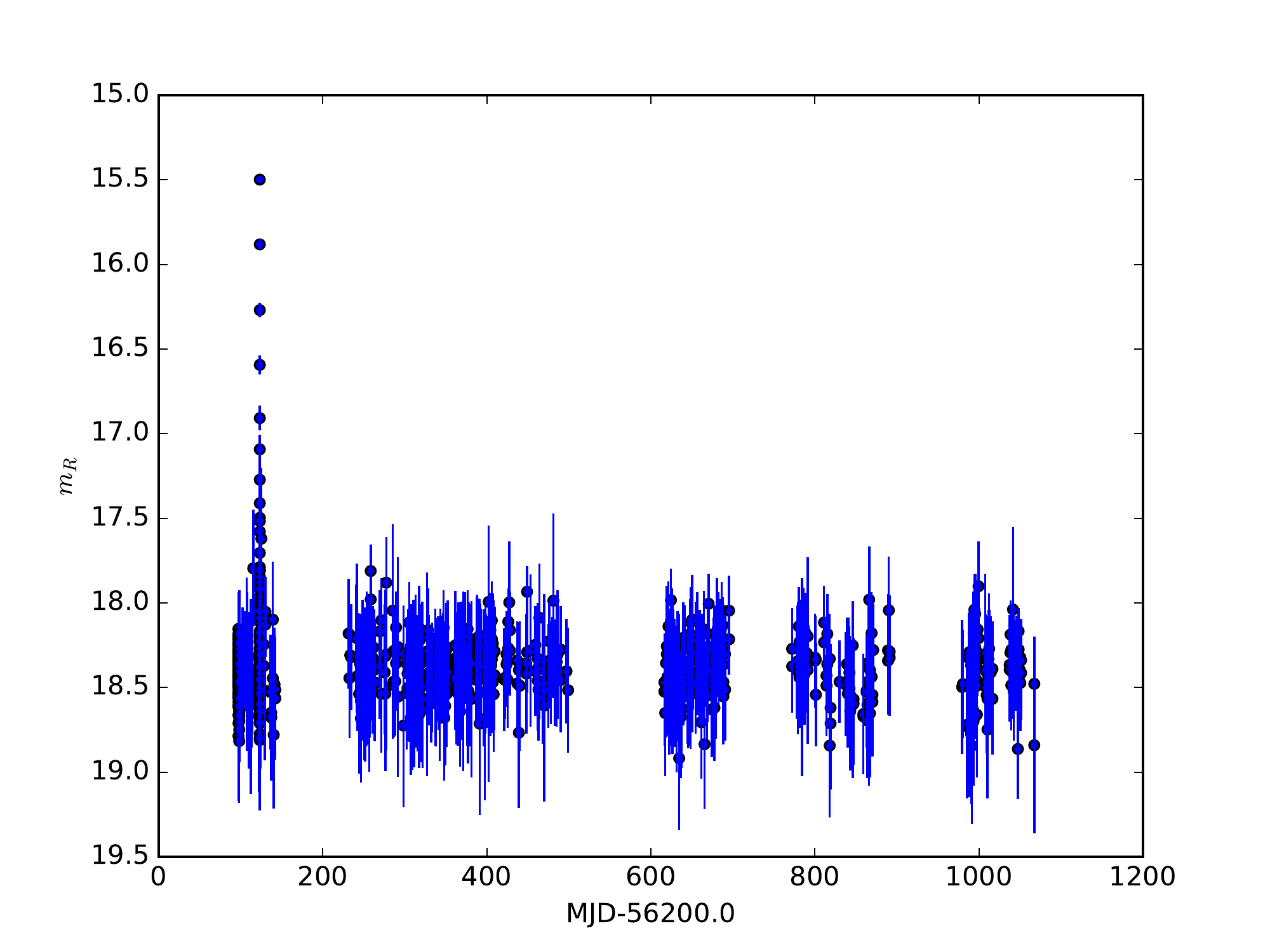}
\caption{{\it Left panel}: Lightcurve of Candidate 86 (or V07979), covering the single night of January 31, 2013 (MJD=56323). The time axis on this plot runs in fractions of a day. {\it Right panel}: Extended lightcurve spanning the whole iPTF observing season of M31 between 2013 and 2015. The lightcurve on the left is a blow-up of the huge excursion in brightness seen around day 150.}\label{fig:86_lc}
\end{figure*}

\subsection{Optical analysis}\label{sec:optical}
As described by SG17, the difference images for the observations are provided by the iPTF pipeline \citep{Masci-2016}, along with a list of objects detected on the difference images, comprising mostly artifacts. 
The total number of \emph{raw detections}, that is those detections obtained from the iPTF pipeline without any characterization of whether they are real or spurious, is $\approx130,000$ for this night.  
We used the method of SG17 to clean these raw detections, based on applying \texttt{WAVDETECT} \citep{Freeman-2002} on the density map of the cumulative raw detections (see SG17 for details), to obtain the list of candidates for variable and transient sources from these observations. It is to be noted that any confirmed transient in this case exhibits variability on a timescale of minutes to hours. We obtained a list of 2232 unique candidates for variable and transient sources.

We then constructed the lightcurves of these candidates via forced aperture photometry performed on the difference images. We used the nova filter presented by SG17, and modified it to pick up fast transients contained within the single-night timescale. In particular, we looked for extreme brightening  events, whereby we increased the multiplicative factor of the threshold in the original algorithm to 10, and decreased the rise-time criterion applied to novae to less than one day (see SG17 for details).  Forty-six candidates passed the filter. We finally examined the surviving candidates by eye. We found that most of them had systematically brightened during the last few epochs of the night, thus pointing to their spurious nature. This is a consequence of the reduced quality of reference-image subtraction toward the end of the night. The corresponding difference images contain halos around many bright stars.

Only one source survived the visual examination, named Candidate 86, whose lightcurve is shown in Fig.~\ref{fig:86_lc}. The source is located at $\mathrm{RA}=00{\rm h}42{\rm m}36.55{\rm s}$ and $\mathrm{DEC}=+41{\rm d}13{\rm m}50.82{\rm s}$ (J2000). It exhibits a 3~mag brightening, and a slower decline with timescale of approximately 2 hours to return close to its quiescent magnitude. It is to be noted that the corresponding quiescent object is obtained from the reference image catalog, and considering Poisson statistics assuming a uniform source density in the crowded bulge of M31, the probability of a chance alignment is $<4\%$ for a search radius of $1''$.
The extended lightcurve of this source, covering the whole period of the iPTF M31 survey between 2013 and 2015, is shown in the right panel of Fig.~\ref{fig:86_lc}; the event of January 31 is not seen to repeat on other nights. Of course, this may be a selection effect as most of the other nights only had two to three epochs. However, two more nights had more than 50 epochs of observations and there was no activity on the source during those nights. This non-detection of another event rules out event rates larger than $2.75\times10^{-3}$ per day at the 95\% confidence level (see Appendix~\ref{append}).

Other optical surveys of M31 with repeat observations (not necessarily high cadence)  have counterparts for Candidate 86, within an error radius of less than $1.5''$. This includes the WeCAPP survey of M31, where \citet{Fliri-2006} reported the source (WeCAPP V07979, hereafter V07979) to be an irregular variable, but with no dramatic variation. To characterize variability of sources they used {\em variation magnitude} defined as  the  difference between the maximum and minimum flux of the source converted to magnitudes, $m(\Delta F)=-2.5\log({\Delta F/F_{\rm vega})}$.\footnote{It is to be noted that the so defined "variation magnitude" is not the same as the amplitude of the variability of the light curve plotted in magnitude units,  $\Delta m=m_{\rm peak}-m_{\rm quiesc}$ (cf.~Fig.~\ref{fig:86_lc}).} The variation magnitude for V07979 determined by \citet{Fliri-2006}  was $m_{R}(\Delta F)\approx 20.0$. Using the data of the   POINT-AGAPE survey of M31,  \citet{An-2004} find the source as a variable star with a variation magnitude of $m_{R}(\Delta F)\approx 20.89$.  For comparison, the flare reported in this work has the variation magnitude of  $m_{R}(\Delta F)\approx 15.5$ (PTF system),  i.e., $\approx 15.3$ in the Vega system \citep{Ofek-2012}. We can thus see that the event observed by iPTF  is unique, exhibiting significantly larger variation than detected in any of the past optical obervations of this source.

\subsection{Optical counterpart in quiescence}

\begin{figure*}
\centering
\includegraphics[width=0.65\textwidth]{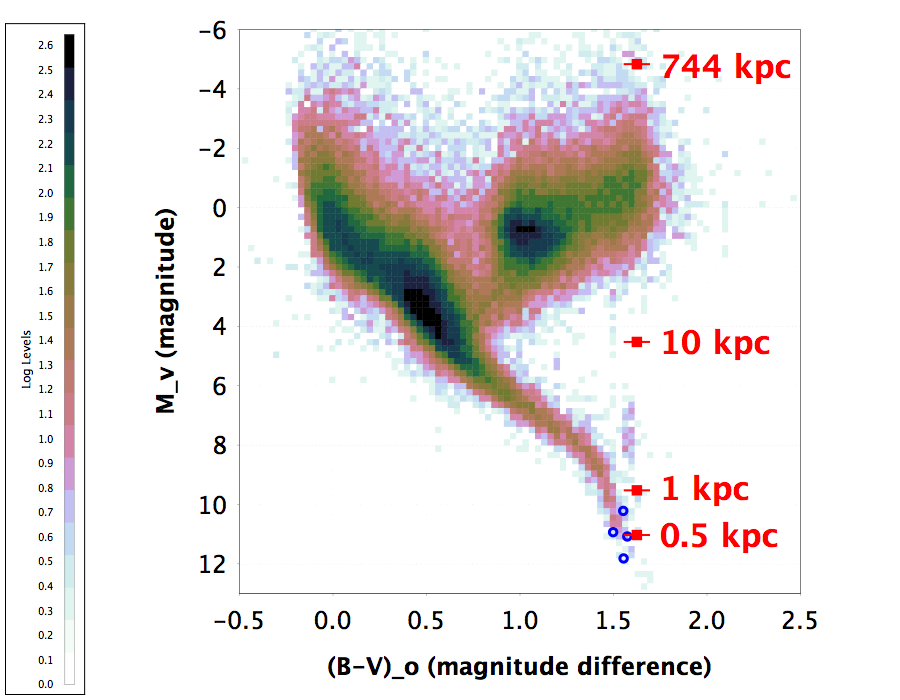}
\caption{Hess diagram showing the location of V07979 depending on the assumed distance, shown by the red squares. The stars used in constructing this Hess diagram are obtained from the the revised Hipparcos catalog  \protect\citep{Leeuwen-2007,Perryman-1997}{\protect\footnotemark}. The error-bars for the V07979 points along the horizontal axis reflect the uncertainty of the reddening correction (see Sect.~\ref{sec:optical}); the error-bars along the vertical axis due to uncertainty in the extinction correction are negligibly small as compared to the axis scale and cannot be seen. The blue circles denote the positions of several well-known UV Ceti stars from \protect\citet{Dal-2010}. }
\label{fig:HR}
\end{figure*}

A number of detections in the optical band have been reported in the literature for this source in quiescence. \citet{Massey-2006} measured $R\approx18.39$ for this star in their imaging of M31 for the Local Group Galaxies Survey (LGGS). This value is consistent with the quiescent magnitude we measure with iPTF (see Fig.~\ref{fig:86_lc}).  
From its quiescent magnitude (in $V$ band) and color from LGGS \citep{Massey-2006}, $V=19.75$ and $B-V=1.69$, V07979, {\em assuming it is located in M31}, belongs to the red supergiant (RSG) branch of the Hertzsprung-Russell (HR) diagram (the red circular point in Fig.~\ref{fig:HR}, where the HR diagram is shown in the Hess form).   As we do not have additional information (for example, spectroscopy) to accurately determine the local extinction at the position of this source, we have considered two cases to localize it in the HR diagram. In the first case, we correct for only foreground extinction using the values $A_{V}=0.20$~mag and $A_{B}=0.25$~mag from \citet{Shafter-2009} obtained using the foreground reddening along the line of sight to M31 from \citet{Schlegel-1998}. It is to be noted that this is the minimal extinction value the source is subjected to, if located in M31. In the second case, we use the 2D extinction map of M31 from \citet{Tempel-2011} and attribute the pixel value at the position of this source (the pixel size of SDSS is roughly equal to the area enclosed by the point spread function in the iPTF images) in the map as the maximal possible {\em intrinsic} extinction for it. The intrinsic extinction values are not drastic, $A_{V,i}=0.05$~mag and $A_{B,i}=0.08$~mag. 
Their sum with the foreground values give us the maximal total extinction the source could suffer.  
The two cases thus lead to the error bars in marking the source's position in the HR diagram.

\citet{Massey-2006} identified two sequences of stars in the two-color diagram of M31 (Figure~11 of their paper), which they claim separate the RSGs in M31 and foreground dwarfs. V07979 (using its colors of  $B-V=1.69$ and $V-R=1.39$ obtained from the LGGS) however does not fall clearly in the compact sequence of foregrounds, and neither does it fall clearly in the loosely bound RSG sequence.

In order to discriminate between different possibilities of the nature of the source, we look for observational information at different wavebands than optical. We find that the source has been detected most often in the X-rays. It has been recorded in the catalogs of X-ray sources produced in the course of many {\it Chandra} and {\it XMM-Newton} observation campaigns of M31. However, a detailed analysis of the X-ray data has never been performed. We pursue in the next section a detailed analysis of the X-ray observations of this source, in particular its spectral analysis to build up a robust multi-wavelength picture throwing light on the classification of this source.

\footnotetext{Compiled by Dr.~E.~Mamajek (\url{http://www.pas.rochester.edu/~emamajek/})}

\setcounter{figure}{2}
\begin{figure*}
\centering
 \includegraphics[width=88mm]{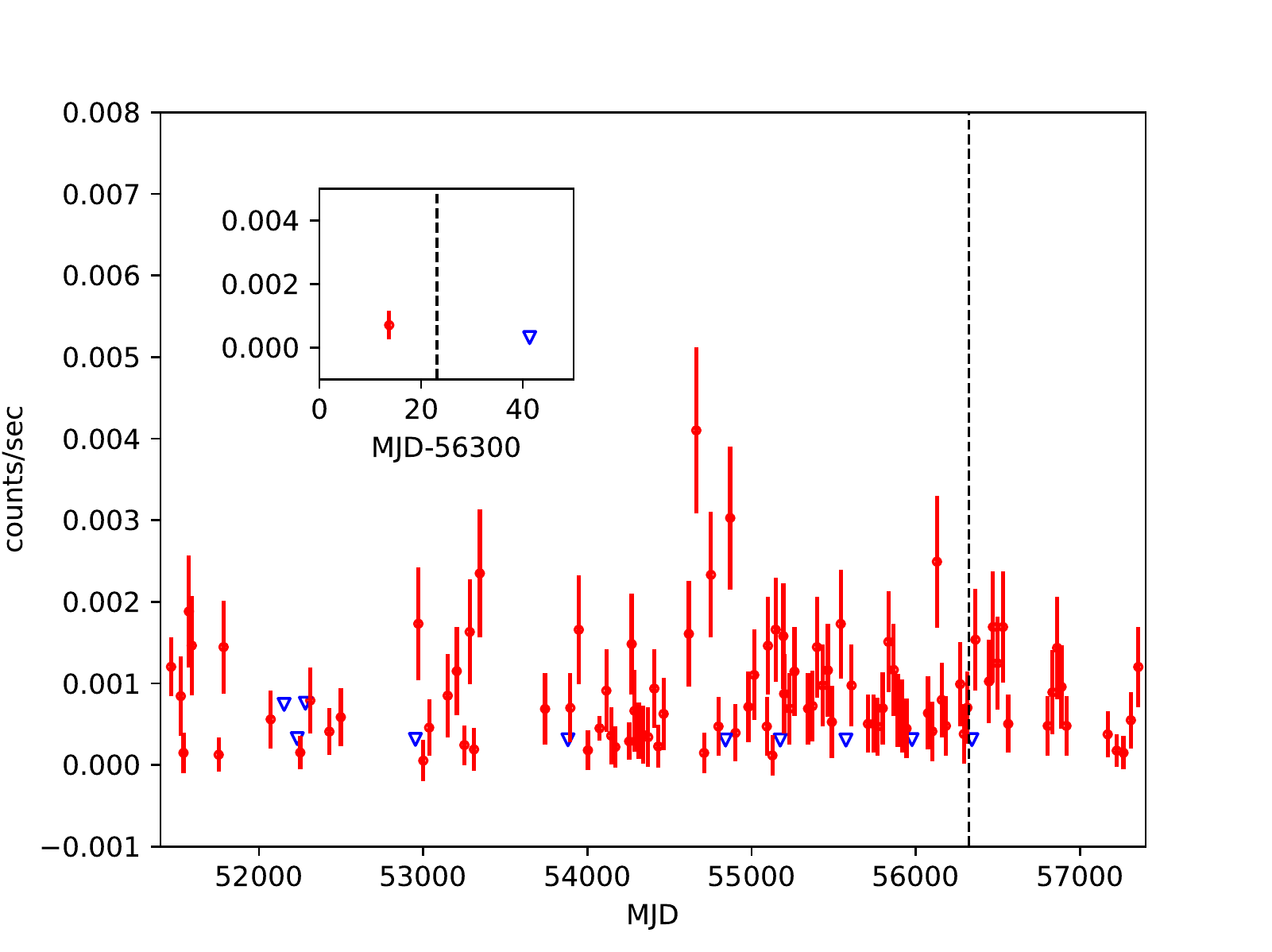}\hfill \includegraphics[width=88mm]{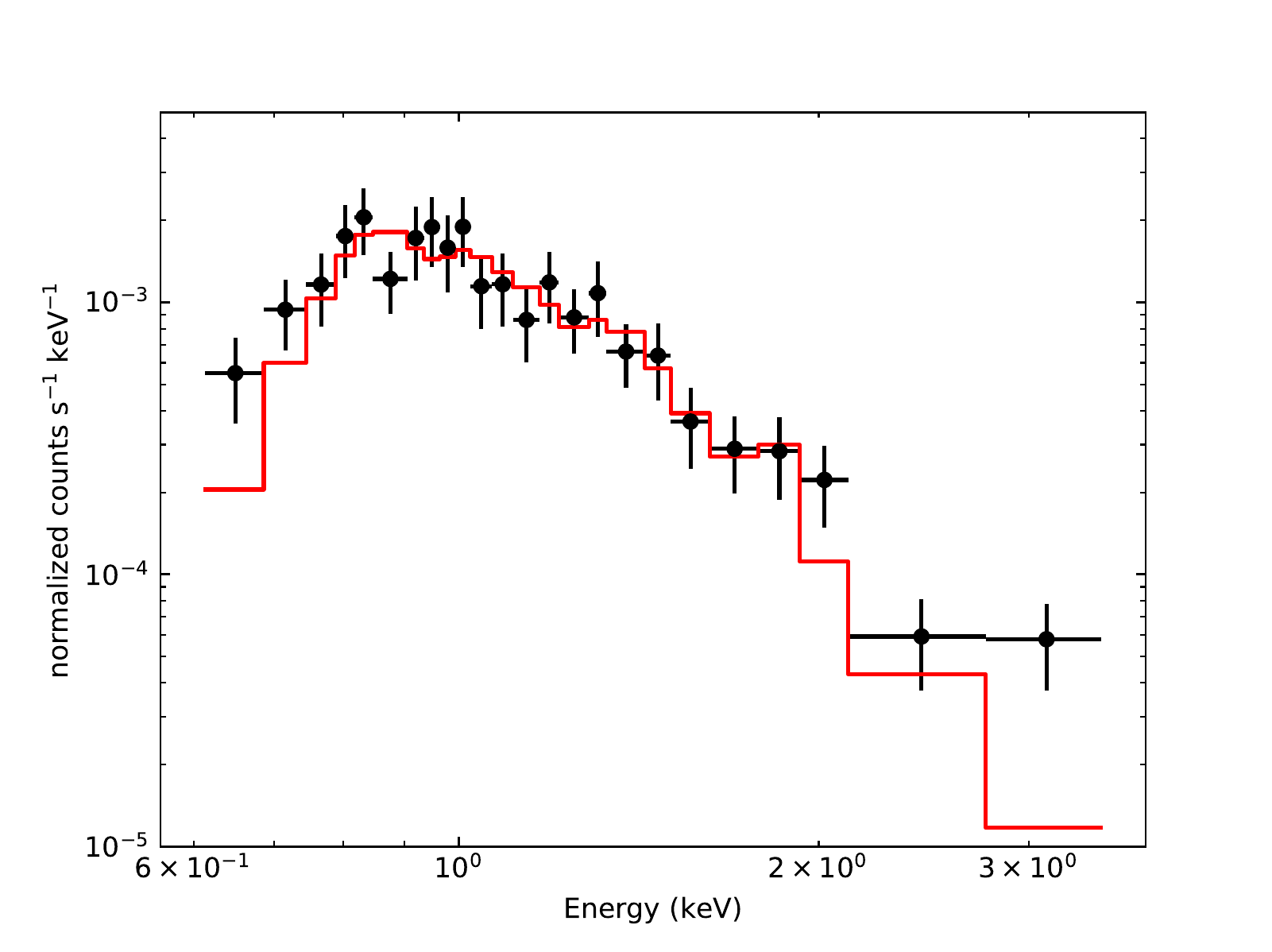}
 \caption{ {\it Left panel}: X-ray lightcurve of V07979 based on the observations from $Chandra$. For those observations where the count rate is zero, upper limits are given, indicated by the blue triangles. Unfortunately, there is no X-ray observation on January 31, 2013, which is marked by the dashed black line, and shown zoomed-in in the inset. {\it Right panel}: Observed spectrum of V07979 extracted from the {\it Chandra} observations in the left panel, where ${\rm S/N}>1.5$ (black points). The red curve is the fitted model (an absorbed thermal plasma model) using \texttt{XSPEC} (see text for details).}
 \label{fig:xray_lc}
\end{figure*}

\section{X-ray data}\label{sec:xray}

In many of the X-ray detections of V07979, the authors have found its X-ray emission to be supersoft or quasi-soft  (for example, \citealt{Stefano-2004, Kong-2002}). Some authors have classified it as a (variable) star (for example, \citealt{Lin-2012, Barnard-2014}) based on the X-ray properties. These authors, however, recommend further investigation, as the value of their characterization parameter (for example, ratio of X-ray to infrared fluxes in the case of \citealt{Lin-2012}) for this source exhibited extreme values.

For our analysis, we gathered all available observations made with the ACIS-I instrument of the $Chandra$ X-ray Observatory that have V07979 in their footprints. 
In total, there are 106 observations for this source with the baseline time covering October 1999 to November 2015 (listed in Table~\ref{table:chandra_obs}). As our source is located in the crowded bulge of M31, we turned to the {\it Chandra} observations given its exquisite spatial resolution.

The {\it Chandra} data are  reduced using \texttt{CIAO} (\texttt{CIAO 4.9}, which is the most recent version) software tools. After reprocessing the data to apply the latest version (by the end of the year 2016) of \texttt{CALDB} (version 4.7.3), we extract the flux (background-subtracted count rate) at the position of V07979 from all the observations using the \texttt{srcflux} routine. We use the energy range from 0.5 to 7.0~keV for the extraction. The resulting X-ray lightcurve is shown in Fig.~\ref{fig:xray_lc} (left panel). 
Even though the baseline covers more than a decade, there is no observation on January 31, 2013, when the optical outburst was detected with iPTF. As can be seen from the lightcurve, V07979 appears to show some level of variability in the X-rays as well, with notable flux excess $\approx 3\mbox{--}4\times10^{-3}$ ($>3\sigma$ of the mean count rate) around MJD 54663 and 54869.

Individual observations do not yield sufficient counts for a spectral analysis. We therefore combined all of the observations where the count rate for V07979 has a significance greater than 1.5 to obtain its spectrum. For this task, we used the routine \texttt{specextract}, and grouped the resulting spectrum using \texttt{grppha}, such that each bin contains more than 10 counts. We then performed the spectral fitting using $\chi^{2}$ statistics. The observed spectrum is shown in the right panel of Fig.~\ref{fig:xray_lc}.

We fitted different models to the observed spectrum using \texttt{Xspec}.  The background was subtracted prior to the model-fitting, as the region around the source may have some diffuse X-rays. 
The best measures of the fitting statistic are obtained for a blackbody spectrum and the spectrum of optically thin plasma (APEC, \citealt{Smith-2001}). The best-fit parameters of both models are summarized in Table~\ref{table:xobs}. The column density is however not constrained with the blackbody model.  As one can see from Table~\ref{table:xobs}, a somewhat better fit to the observed spectrum is  provided by the APEC model. This model is plotted, along with the observed spectrum, in Fig.~\ref{fig:xray_lc}. For the interstellar absorption, we used the {\it tbabs} model with the ISM abundances from \citet{Wilms-2000}.

\begin{table*}
\centering
\caption{Best-fit parameter values of the X-ray spectral models.}
\label{table:xobs}
\renewcommand\arraystretch{1.5}
\begin{tabular}[width=\textwidth]{c | c | c}
\hline
Parameter name  &APEC model  &Blackbody model\\
\hline
Hydrogen column density, $N_{\rm H} ({\rm cm}^{-2})$         &$7.16(\pm 0.9)\times 10^{21}$      &$3.07\times 10^{13}$ (unconstrained){\tablefootmark{a}}\\
Temperature, $kT ({\rm keV})$                      &$0.61(\pm 0.06)$                                              &$0.23(\pm 0.01)$\\
Reduced chi-squared,  $\chi^{2}_{\rm red}$      &1.041 (21 d.o.f)                                      &1.333 (21 d.o.f)\\
\hline
Absorbed flux in 0.5-7.0~keV range (erg~${\rm cm}^{-2}$~${\rm s}^{-1}$)                      			&$9.89\times10^{-15}$               				&$1.12\times 10^{-14}$\\
Unabsorbed flux in 0.5-7.0~keV range (erg~${\rm cm}^{-2}$~${\rm s}^{-1}$)                      			&$4.36\times 10^{-14}$               				&$1.12\times 10^{-14}$
\end{tabular}
\tablefoot{\tablefoottext{a}{This value of the column density for the blackbody model is effectively zero, and thus the absorbed and unabsorbed fluxes for this model are the same.}}
\end{table*}

\begin{figure}
\centering
\includegraphics[width=88mm]{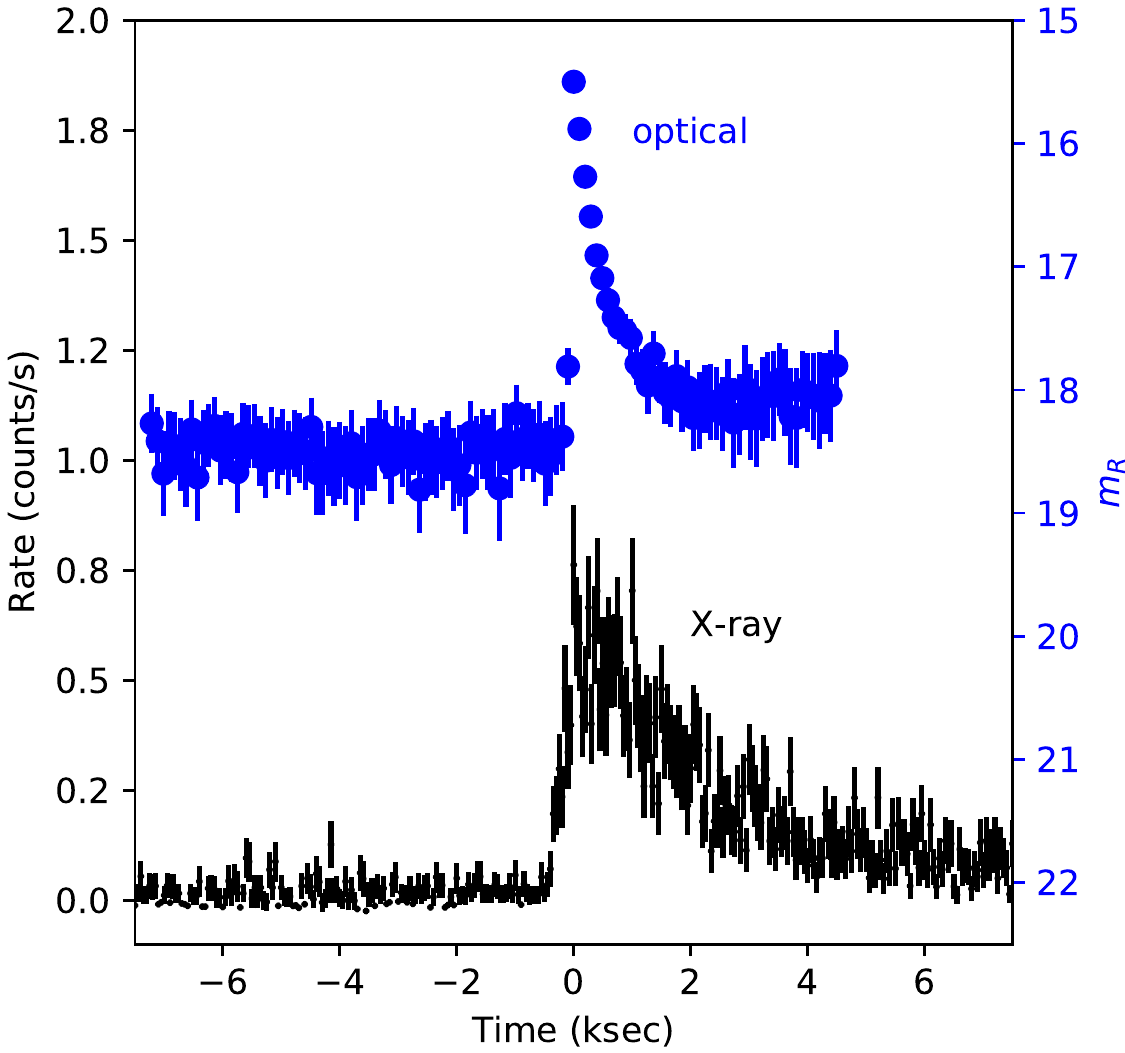}
\caption{({\it top}, {\it blue}) optical iPTF lightcurve of V07979 from January 31, 2013 (Fig.~\ref{fig:86_lc} left panel), and ({\it bottom}, {\it black}) X-ray flare at the position of V07979 from an {\it XMM-Newton} observation of the central part of M31 on January 28, 2008 produced by the PPS of {\it XMM-Newton} SSC. The horizontal axis shows time with respect to the corresponding peak time for both lightcurves.}\label{fig:xflare}
\end{figure}

\section{Nature of the source}\label{sec:result}

There is a well-known ambiguity in classification of stars of  late spectral types located in the direction of M31. Due to the value of the M31 distance modulus of $\mu\sim 24$ \citep{Vilardell}, stars with $B-V\sim 1.5$ located in this part of the sky can be either bright red super-giants located in M31 or much less luminous low-mass main sequence stars located in the  disk of the Milky Way, within $\sim 0.5\mbox{--}1$~kpc from the Sun. In the absence of spectroscopic data, this ambiguity is often difficult to solve based solely on the optical data (although the two-color diagram analysis may be of certain help, cf.~\citet{Massey-2006}). This  complicates the classification of V07979, as illustrated by Fig.~\ref{fig:HR}.

If we proceed with considering V07979 as a RSG, this appears to preclude the January 31 outburst to be related to chromospheric activity, as this is generally considered to be absent in late supergiant stars -- the rotational energy needed for the dynamo regeneration of magnetic fields gets dissipated via their strong stellar winds \citep{Guedel}.    
As such, for this transient in M31, its kilo-second timescale and the amount of radiation energy involved makes it rather unique and fascinating on first glance.  
The source is not known to exhibit any such activity in the optical as observed with the iPTF, in the earlier literature. Even the possibility of it being the first optical flare seen in a supergiant star poses an interesting avenue to investigate.  Another exciting possibility would be that the observed flare is a very fast nova explosion, in which case it would have to occur on a very massive white dwarf, very close to the Chandrasekhar mass limit, accreting mass at a high rate. Also, a very short recurrence time ($\lesssim$ a few years) should be expected in this case, which is not ruled out by the upper limit on the event rate of V07979 (Appendix~\ref{append}).

From results of X-ray spectroscopy, we see that although the source is  soft -- the blackbody fit yields a temperature of $kT_{\rm eff}\approx 0.23$ keV, but it is not supersoft, defined as $kT_{\rm eff}\lesssim 0.1\mbox{--}0.15$~keV. This  rules out the possibility of V07979 harboring a nuclear burning accreting WD, as such sources are characterized by a supersoft spectrum, with $kT\lesssim 150$ eV even for the most massive white dwarfs near the Chandrasekhar mass limit \citep{Kahabka, Wolf, Gilfanov-2010}. However, the X-ray spectra of novae during quiescence are harder than during the supersoft phase \citep[e.g.,][]{Zemko-2015}. On the other hand, the fact that the spectrum is well approximated with the optically thin plasma emission model and the value of its best-fit temperature, $kT_{\rm APEC}\approx 0.6$ keV, are consistent with its interpretation as coronal emission from a star or an active binary. 

The average (unabsorbed) flux from the APEC model (Table~\ref{table:xobs}) corresponds to a 0.5-7.0~keV luminosity  of $3\times10^{36}$~erg/s, assuming that the source is located in M31. This is obviously too large for the coronal emission. Indeed, maximal X-ray luminosities  emitted by active (including young) stars and binaries such as RS CVn systems are typically in the range $\sim 10^{29}\mbox{--}10^{32}$~erg/s \citep{Seward-1995,Feigelson-2007,Pye-2015}. Such luminosities are also too large for colliding wind binaries, which rarely exceed $10^{35}$ erg/s \citep[e.g.,][]{Gagne-2012}.  Furthermore, X-ray emission from novae at quiescence typically produces a few times $10^{30}$~erg/s \citep{Orio-2001}, which does not come anywhere close to the inferred luminosity of V07979 at the distance of M31.
An X-ray luminosity on the order of $\sim 10^{36}$ erg/s can be easily produced by an X-ray binary, however, at these luminosities, spectra of accreting black holes and neutron stars are much harder than observed for V07979 \citep{Gilfanov-2010LNP}.

The optical and X-ray data can be reconciled assuming that the source is located in the disk of the Milky Way, within $\sim 0.5\mbox{--}1$~kpc from the Sun. Its location in the HR diagram assuming a distance of 10, 1 and 0.5~kpc is shown in Fig.~\ref{fig:HR}.  In the same figure, the positions of archetypical UV Ceti-type stars from \citet{Dal-2010} are also shown. UV Ceti-type stars are flare stars containing red (spectral type M) dwarfs, in which the amplitude of the flare can reach several magnitudes. For an assumed distance of 0.5--1~kpc, V07979 sits on the red dwarf main sequence belt, where the other UV Ceti-type stars are also located. Its coronal temperature of $\approx 7$~MK and its resulting X-ray luminosity (0.5-7~keV) of $\approx 1.3\mbox{--}5\times10^{30}$~erg/s, are consistent with that of active/flare stars, for example the prototype UV Ceti stars \citep{Audard-2003} and the solar neighborhood flare stars \citep{Pallavicini-1990}. The ratio of its X-ray flux to quiescent optical $R$ band flux measures $\approx -1.2$~dex and this value is in agreement with that of flaring stars $\approx -0.9\pm1.5$~dex deduced from \citet{Lin-2012}.

Thus, all these pieces of evidence point to the January 31, 2013 event of V07979 (Fig.~\ref{fig:86_lc}) being a coronal flare from an M-type main-sequence star in the Milky Way.
The flares of UV Ceti-type stars are most intense at shorter wavelengths, given that they arise from very hot plasma heated by the release of magnetic energy. A peak flare amplitude of $\gtrsim 3$~mag in the red $R$ band is hard to observe for these objects. Such large amplitudes have been seen only in a few cases (for example, \citealt{Chang-2015}). Thus, this discovery of a large-amplitude flare on V07979 with iPTF is quite remarkable in itself.

The flare nature of the optical event is further confirmed by the detection of an X-ray flare at the position of our source in archival {\it XMM-Newton} data. Figure~\ref{fig:xflare} shows the light curve of this X-ray flare, which was observed on January 28 of 2008 ({\it XMM-Newton} obsID 0505720501, PI Wolfgang Pietsch) and obtained from the Pipeline Processing System (PPS) data product of {\it XMM-Newton} Survey Science Center (SSC). Measurements related to this flare were reported by \citet{Lin-2012}, but without conclusive localization of the source. 
The similarity between this lightcurve and the optical lightcurve (shown in the same figure in blue), including the duration, even though not contemporaneous, is striking.

\subsection{Consequences of the coronal flare scenario}

We estimate the main parameters of the flare from V07979. Following \citet{Chang-2015}, for this event we find a rise time of $\approx 3.5$~mins and a decay time of $\approx 75$~mins. Its {\em equivalent duration} computed using their Equation~7 is $\approx 95$~mins. Assuming a distance of 0.5-1~kpc, we obtain its peak luminosity in the optical band of $\approx 0.7\mbox{--}3\times 10^{32}$~erg/s and its flare energy (via their Equation~8) of $\approx 0.3\mbox{--}1\times 10^{35}$~erg. Thus, with its energy exceeding $10^{33}$~erg, this event of V07979 can be classified as a superflare \citep[e.g.][]{Shibayama-2013, Chang-2015}.  

The rise and decay times of the V07979 flare are consistent with the distributions obtained by \citet{Chang-2015}. However, with the flare energy of $>10^{34}$~erg, this event is among the most energetic  flares observed from M-stars. The upper limit on the frequency of such flares from this object is derived in Appendix~\ref{append} based on the non-detection of other events of this magnitude during the iPTF observations performed in 2013--2015. It equals $\log f ({\rm hr}^{-1})\lesssim-3.94$. Strong stellar flares occur much less frequently than weak ones, with the typical frequency of events of such a large magnitude as the V07979 flare being $\log f ({\rm hr}^{-1})\sim -3\mbox{--}-4$ \citep[e.g.,][]{Chang-2015}. Our upper limit on its rate is thus consistent with the energetics we have found above.

\section{Discussion and Conclusions}\label{sec:conclude}

A very fast and  bright transient event, associated with V07979,  was discovered in the bulge of M31 during iPTF high cadence observations of this galaxy on January 31, 2013. The optical counterpart of this event in quiescence was thought to be a star located in M31. There are a number of exciting possibilities if this event were located in M31. However, we argue that  the V07979 event was a stellar superflare on a main-sequence M-type star located  in the Milky Way along the line of sight to M31, at a distance of $\approx 0.5-1$ kpc from the Sun. 
Even though the persistent counterpart of  this event has been registered in a number of  optical and X-ray surveys of M31, its definitive characterization and localization, through a multi-wavelength analysis is presented for the first time here. We argue that it is a UV Ceti type dwarf star with the quiescent magnitude of $M_R\approx 11.0\mbox{--}9.5$  and average quiescent X-ray luminosity of  $\approx 10^{30}$~erg/s. During iPTF observations it produced a superflare of amplitude around 3~mag in the red filter of iPTF, which is rare to observe in stellar flares. The event had an equivalent duration of around 95~mins and optical energy  of $\approx (0.3\mbox{--}1)\times10^{35}$~erg. The future data releases of Gaia could potentially give other useful information of the source such as proper motion, that was not present in the first data release.

Our approach of using multi-wavelength observations of a transient object, as well as the specific considerations detailed in Sect.~\ref{sec:result}, can be seen as an example to follow in future classification attempts of transient or variable objects. This will be especially useful for the purpose of extracting the rarest and most energetic extragalactic events from the quickly accumulating time-domain detections from upcoming surveys, greatly reducing the risk of confusing these with less energetic foreground events.

This work also serves as another demonstration of the varied science case possibilities with time-domain data. The use of particularly high-cadence (minutes) optical time-domain data, even if restricted by a single filter, when complemented with observations from a different wavelength regime, can lead to the discovery of interesting kinds of astrophysical objects. The iPTF survey, with its varied-cadence experiments, has provided us with this interesting object from its M31 observations. We presently continue the hunt for other types of transients in different fields of M31.

\section*{Software}
\texttt{DAOPHOT} \citep{Stetson}, \texttt{DAOGROW} \citep{Stetson-1990}, \texttt{HEASoft}, \texttt{CIAO} \citep{CIAO-2006}, \texttt{numpy} \citep{numpy}, \texttt{scipy} \citep{scipy}, \texttt{astropy} \citep{astropy}, \texttt{pandas} \citep{pandas}, \texttt{mpi4py} \citep{mpi4py}, \texttt{matplotlib} \citep{matplotlib}.

\section*{Acknowledgments}
This research has made use of data obtained from the Chandra Data Archive and the Chandra Source Catalog, and software provided by the Chandra X-ray Center (CXC) in the application packages \texttt{CIAO}, \texttt{ChIPS}, and \texttt{Sherpa}. It is also based on observations obtained with XMM-Newton, an ESA science mission with instruments and contributions directly funded by ESA Member States and NASA. 
MG acknowledges partial support by Russian Scientific Foundation (RNF), project 14-22-00271.
This research has made use of software obtained from the High Energy Astrophysics Science Archive Research Center (HEASARC), a service of the Astrophysics Science Division at NASA/GSFC and of the Smithsonian Astrophysical Observatory's High Energy Astrophysics Division. We also thank the anonymous referee for the many helpful comments and suggestions that helped improve this paper.
\bibliographystyle{aa}
\bibliography{ref_flare}

\appendix
\section{Upper limit on the frequency of outbursts from V07979}\label{append}

\begin{figure}
 \includegraphics[width=88mm]{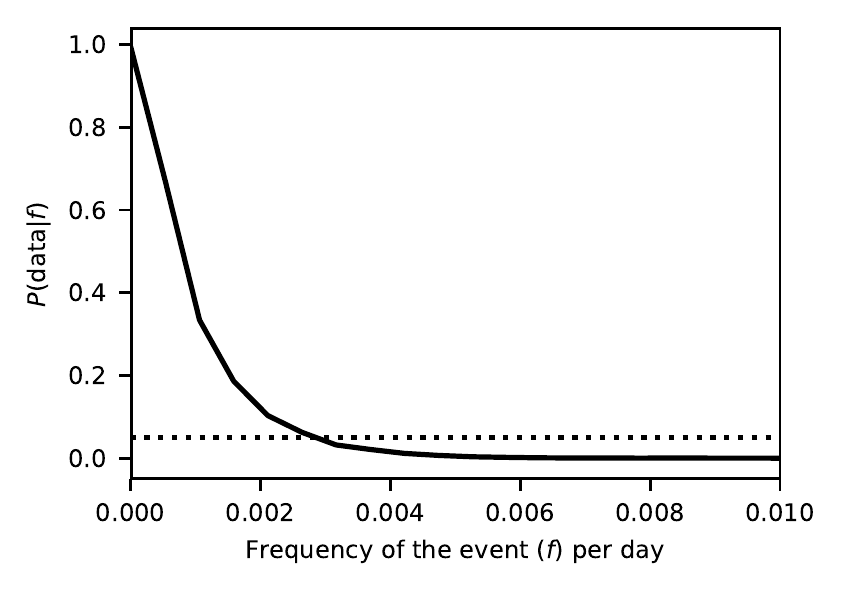}
 \caption{Probability of not observing another transient event on V07979 as a function of assumed event rate. The doted horizontal line marks the 5\% level.} 
 \label{fig:rate}
 \end{figure}

We can use the absence of {\em any trace} of a second similar event during the iPTF observations (Sect.~\ref{sec:optical}) to deduce an upper limit on the repetition rate for such events. For this, we assume that the event repeats in a random way following Poissonian statistics with a fixed rate $f$. We run a series of Monte-Carlo simulations to assess the probability of detecting a second event with iPTF as a function of the rate. In each simulation we fix one event to occur at the time of the event detected in the real observations and place other events in its future and past by drawing their time differences independently from an exponential distribution, which is the distribution of intervals between events for Poisson-distributed events with the given rate. We further assume that the outburst profile  is same for all simulated events, both in shape and maximum magnitude (see Fig.~\ref{fig:86_lc} left panel). For the relevant values of the rate $f$, the finite duration of this profile is much shorter than the intervals between events and therefore it  does not matter for the placement of the simulated events in time.
We then calculate the probability of not detecting a second event, using the iPTF observing pattern and a detection threshold of $m_{R}=18$, by evaluating 10000 random simulations per rate.  
Figure~\ref{fig:rate} shows this probability as a function of the assumed rate, from which we can read off an upper limit on the rate at the 5\% probability (corresponding to the 95\% confidence level) of $2.75\times10^{-3}$ events per day.\footnote{The interpretation of this upper limit is that any higher rate would yield a second observed event with a likelihood of more than 95\%.}

\begin{table*}
\centering
\caption{\emph{Chandra} observations of M31 used in this paper.}
\label{table:chandra_obs}
\renewcommand\arraystretch{1.0}
\begin{tabular}[width=\textwidth]{l l c l l | l l c l l}
\hline
ObsID  &Inst  &Exp (ks)  &Obsdate  &PI  &ObsID  &Inst  &Exp (ks)  &Obsdate  &PI\\
\hline
303    &ACIS-I   &11.8   &1999-10-13   &Murray       &10715    &ACIS-I   &4.0   &2009-09-18   &Murray\\
305    &ACIS-I   &4.1   &1999-12-11   &Murray        &10716    &ACIS-I   &4.1   &2009-09-25   &Murray\\
306    &ACIS-I   &4.1   &1999-12-27   &Murray        &10717    &ACIS-I   &4.1   &2009-10-22   &Murray\\
307    &ACIS-I   &4.1   &2000-01-29   &Murray        &10719    &ACIS-I   &4.1   &2009-12-27   &Murray\\
308    &ACIS-I   &4.0   &2000-02-16   &Murray        &11275    &ACIS-I   &4.1   &2009-11-11   &Garcia\\
311    &ACIS-I   &4.9   &2000-07-29   &Murray        &11276    &ACIS-I   &4.0   &2009-12-08   &Garcia\\
312    &ACIS-I   &4.7   &2000-08-27   &Murray        &11277    &ACIS-I   &4.1   &2010-01-01   &Garcia\\
1577    &ACIS-I   &4.9   &2001-08-31   &Murray       &11278    &ACIS-I   &4.0   &2010-02-04   &Garcia\\
1583    &ACIS-I   &4.9   &2001-06-10   &Murray       &11279    &ACIS-I   &4.1   &2010-03-05   &Garcia\\
1585    &ACIS-I   &4.9   &2001-11-19   &Murray       &11838    &ACIS-I   &4.0   &2010-05-27   &Murray\\
2895    &ACIS-I   &4.9   &2001-12-07   &Garcia       &11839    &ACIS-I   &4.0   &2010-06-23   &Murray\\
2896    &ACIS-I   &4.9   &2002-02-06   &Garcia       &11840    &ACIS-I   &4.0   &2010-07-20   &Murray\\
2897    &ACIS-I   &4.9   &2002-01-08   &Garcia       &11841    &ACIS-I   &4.0   &2010-08-24   &Murray\\
2898    &ACIS-I   &4.9   &2002-06-02   &Garcia       &11842    &ACIS-I   &4.0   &2010-09-25   &Murray\\
4360    &ACIS-I   &4.9   &2002-08-11   &Primini      &12160    &ACIS-I   &4.0   &2010-10-19   &Murray\\
4678    &ACIS-I   &3.9   &2003-11-09   &Murray       &12162    &ACIS-I   &4.0   &2010-12-12   &Murray\\
4679    &ACIS-I   &3.8   &2003-11-26   &Murray       &12163    &ACIS-I   &4.0   &2011-01-13   &Murray\\
4680    &ACIS-I   &4.2   &2003-12-27   &Murray       &12164    &ACIS-I   &4.0   &2011-02-16   &Murray\\
4681    &ACIS-I   &4.1   &2004-01-31   &Murray       &12970    &ACIS-I   &4.0   &2011-05-27   &Garcia\\
4682    &ACIS-I   &4.0   &2004-05-23   &Murray       &12971    &ACIS-I   &4.0   &2011-06-30   &Garcia\\
4719    &ACIS-I   &4.1   &2004-07-17   &Garcia       &12972    &ACIS-I   &4.0   &2011-07-25   &Garcia\\
4720    &ACIS-I   &4.1   &2004-09-02   &Garcia       &12973    &ACIS-I   &3.9   &2011-08-25   &Garcia\\
4721    &ACIS-I   &4.1   &2004-10-04   &Garcia       &12974    &ACIS-I   &4.0   &2011-09-28   &Garcia\\
4722    &ACIS-I   &3.9   &2004-10-31   &Garcia       &13298    &ACIS-I   &3.9   &2012-05-26   &Murray\\
4723    &ACIS-I   &4.0   &2004-12-05   &Garcia       &13299    &ACIS-I   &3.9   &2012-06-21   &Murray\\
7064    &ACIS-I   &23.2   &2006-12-04   &Murray      &13300    &ACIS-I   &3.9   &2012-07-20   &Murray\\
7068    &ACIS-I   &7.7   &2007-06-02   &Murray       &13301    &ACIS-I   &3.8   &2012-08-19   &Murray\\
7136    &ACIS-I   &4.0   &2006-01-06   &Garcia       &13302    &ACIS-I   &3.9   &2012-09-12   &Murray\\
7137    &ACIS-I   &4.0   &2006-05-26   &Garcia       &13833    &ACIS-I   &4.0   &2011-10-31   &Garcia\\
7138    &ACIS-I   &4.1   &2006-06-09   &Garcia       &13834    &ACIS-I   &3.9   &2011-11-24   &Garcia\\
7139    &ACIS-I   &4.0   &2006-07-31   &Garcia       &13835    &ACIS-I   &3.9   &2011-12-19   &Garcia\\
7140    &ACIS-I   &4.1   &2006-09-24   &Garcia       &13836    &ACIS-I   &3.9   &2012-01-16   &Garcia\\
8183    &ACIS-I   &4.0   &2007-01-14   &Murray       &13837    &ACIS-I   &3.9   &2012-02-19   &Garcia\\
8184    &ACIS-I   &4.1   &2007-02-14   &Murray       &14927    &ACIS-I   &3.9   &2012-12-09   &Barnard\\
8185    &ACIS-I   &4.0   &2007-03-10   &Murray       &14928    &ACIS-I   &3.9   &2012-12-31   &Barnard\\
8186    &ACIS-I   &4.1   &2007-11-03   &Murray       &14929    &ACIS-I   &3.9   &2013-01-21   &Barnard\\
8187    &ACIS-I   &3.8   &2007-11-27   &Murray       &14930    &ACIS-I   &3.9   &2013-02-18   &Barnard\\
8191    &ACIS-I   &4.0   &2007-06-18   &Garcia       &14931    &ACIS-I   &3.9   &2013-03-12   &Barnard\\
8192    &ACIS-I   &4.1   &2007-07-05   &Garcia       &15324    &ACIS-I   &3.9   &2013-06-02   &Murray\\
8193    &ACIS-I   &4.1   &2007-07-31   &Garcia       &15325    &ACIS-I   &3.9   &2013-06-26   &Murray\\
8194    &ACIS-I   &4.0   &2007-08-28   &Garcia       &15326    &ACIS-I   &3.9   &2013-07-24   &Murray\\
8195    &ACIS-I   &4.0   &2007-09-26   &Garcia       &15327    &ACIS-I   &3.9   &2013-08-26   &Murray\\
9520    &ACIS-I   &4.0   &2007-12-29   &Murray       &15328    &ACIS-I   &4.0   &2013-09-27   &Murray\\
9521    &ACIS-I   &4.0   &2008-11-27   &Murray       &16294    &ACIS-I   &3.9   &2014-05-26   &Murray\\
9522    &ACIS-I   &4.0   &2008-07-15   &Murray       &16295    &ACIS-I   &3.9   &2014-06-24   &Murray\\
9523    &ACIS-I   &4.1   &2008-09-01   &Murray       &16296    &ACIS-I   &3.9   &2014-07-22   &Murray\\
9524    &ACIS-I   &4.1   &2008-10-13   &Murray       &16297    &ACIS-I   &3.9   &2014-08-18   &Murray\\
9529    &ACIS-I   &4.1   &2008-05-31   &Garcia       &16298    &ACIS-I   &3.9   &2014-09-16   &Murray\\
10551    &ACIS-I   &4.0   &2009-01-09   &Garcia      &17443    &ACIS-I   &5.0   &2015-05-26   &Murray\\
10552    &ACIS-I   &4.0   &2009-02-07   &Garcia      &17444    &ACIS-I   &5.0   &2015-07-20   &Murray\\
10553    &ACIS-I   &4.1   &2009-03-11   &Garcia      &17445    &ACIS-I   &5.0   &2015-08-30   &Murray\\
10554    &ACIS-I   &4.0   &2009-05-29   &Garcia      &17446    &ACIS-I   &5.0   &2015-10-14   &Murray\\
10555    &ACIS-I   &4.1   &2009-07-03   &Garcia      &17447    &ACIS-I   &5.0   &2015-11-28   &Murray\\
\hline
\end{tabular}
\tablefoot{The observation ID, instrument used, exposure time, observation date and corresponding principal investigator are indicated by the columns ObsID, Inst, Exp, Obsdate and PI, respectively.}
\end{table*}

\end{document}